\documentclass{biometrika}

\usepackage{amsfonts}
\newcommand*{\eg}{\textit{e.g.}}

\newcommand{\D}[2][]{\frac{\mathrm{d} #1}{\mathrm{d} #2}} 
\newcommand{\di}{\mathrm{d}}

\DeclareMathOperator{\tr}{tr}

\newcommand{\bigmid}{\bigm\vert}

\newcommand{\indep}{\mathbin{\perp\!\!\!\perp}}
\newcommand{\vindep}{\mathbin{\,\ddagger\,}}
\newcommand\bigsym[1]{
\operatornamewithlimits{%
  \mathchoice{\vcenter{\hbox{\huge $#1$}}}
             {\vcenter{\hbox{\Large $#1$}}}
             {\mathrm{#1}}
             {\mathrm{#1}}}}
\newcommand*\bigindep{\bigsym{\indep}}

\newcommand{\law}{\pounds} 
\newcommand{\ddirich}{\mathcal{D}}
 
\newcommand{\dnorm}{\mathcal{N}} 

\newcommand{\R}{\mathbb{R}}

\newcommand{\rtheta}{\widetilde{\theta}}
\newcommand{\rlambda}{\widetilde{\lambda}}

\newcommand{\ralpha}{\widetilde{\alpha}}
\newcommand{\rbeta}{\widetilde{\beta}}
\newcommand{\reta}{\widetilde{\eta}}

\newcommand{\lasso}{\mbox{\sc lasso}}
\renewcommand{\emph}{}
\renewcommand{\top}{{\mbox{\scriptsize\rm T}}}

\accessdate{Accepted for publication, 1st September 2013}

\usepackage{times}
\usepackage{bm}
\usepackage{natbib}

\title{Retrospective--prospective symmetry in the likelihood and
  Bayesian analysis of case-control studies}

\author{Simon P.~J.~Byrne}
\affil{Department of Statistical Science, %
  University College, London WC1E 6BT, U.K. %
  \email{simon.byrne@ucl.ac.uk}}
\author{\and A.~Philip~Dawid}
\affil{Statistical Laboratory, University of Cambridge, %
  Wilberforce Road, Cambridge CB3 0WB, U.K. %
  \email{apd@statslab.cam.ac.uk}}

\begin{document}
\maketitle

\begin{abstract}
  \citet{prentice1979} established that the maximum likelihood
  estimate of an odds-ratio in a case-control study is the same as
  would be found by fitting a logistic regression: in other words, for
  this specific target the incorrect prospective model is
  inferentially equivalent to the correct retrospective model.
  Similar results have been obtained for other models, and conditions
  have also been identified under which the corresponding Bayesian
  property holds, namely that the posterior distribution of the
  odds-ratio be the same, whether computed using the prospective or
  the retrospective likelihood.  Here we demonstrate how these results
  follow directly from certain parameter independence properties of
  the models and priors, and identify prior laws that support such
  reverse analysis, for both standard and stratified designs.
\end{abstract}


In order to estimate the effects of risk factors on a binary outcome,
for example a disease, there are two basic experimental approaches: a
\emph{prospective} or \emph{cohort} study, in which subjects are
selected from the population, possibly based on their risk factors,
and observed to determine if the disease arises; and a
\emph{case-control} or \emph{retrospective} study, in which random
samples are taken from both the subpopulation with the disease, the
cases, and the subpopulation without, the controls, and the
relative frequencies of the risk factors in the two samples are
recorded.  Case-control studies are often desirable or unavoidable,
particularly where the disease is relatively rare or the time to
diagnosis is long, since the costs of obtaining a sufficient sample
size for a prospective study are then likely to be prohibitive.

Let $Y$ be the outcome variable, taking values coded $0$ or $1$,
corresponding to the absence or presence of disease, respectively.
Let $X$ be the vector of covariates, or risk factors, taking values in
$\mathcal{X} \subseteq \R^k$.  In a prospective study we are sampling
from the conditional distribution of $Y$ given $X$.  Under a
proportional odds assumption, the model is that of a logistic
regression:
\begin{equation}
  \label{eq:logistic-dens}
  p(y \mid x,\alpha,\beta) 
  = \frac{e^{y(\alpha+\beta^\top x)}}{1+e^{\alpha+\beta^\top x}},
  \quad \alpha \in \R, \beta \in \R^k.
\end{equation}

A case-control study, however, will result in observations generated from the
conditional distribution of $X$ given $Y$.  In this case, specifying and
analysing the probabilistic model become much more difficult, particularly if
$\mathcal{X}$ is large or infinite.  But \citet{prentice1979} showed that the
maximum likelihood estimator of the log odds-ratio parameter $\beta$, as well
as its asymptotic covariance matrix, can be computed from a logistic
regression: in other words, we can use the incorrect but simpler prospective
model to analyse data gathered retrospectively.  This result has been widely
applied in epidemiology and other areas.  Other models have since been
identified that satisfy this property, notably the multinomial logistic
\citep{baker1994}, the stereotype model \citep{greenland1994}, and the
multiplicative intercept model \citep{weinberg1993}.

There exist analogous results for Bayesian analysis, showing that, for
an appropriately chosen prior distribution, the posterior distribution
of $\beta$ can be computed using the incorrect prospective likelihood
instead of the true retrospective likelihood.  \citet{zelen1986},
\citet{nurminen1987}, \citet{marshall1988} and \citet{ashby1993}
developed this analysis for the case of a single binary covariate:
this involves computing the posterior distribution of the log
odds-ratio of a $2 \times 2$ contingency table under a Dirichlet
prior.  For the case of categorical covariates, where $\mathcal{X}$ is
finite, \citet{seaman2004} identified a class of improper priors that
satisfy the desired properties; this class was extended to include
proper priors by \citet{staicu2010}.  Extensions to stratified and
general multinomial designs have been studied by
\citet{ghosh2006,ghosh2012}.

With the advent of computational tools such as Markov chain Monte
Carlo simulation, direct analysis of the retrospective likelihood need no longer
present an obstacle.  \citet{muller1997}, \citet{seaman2001} and
\citet{gustafson2002} have pursued this approach, which is reviewed in
\citet{mukherjee2005}.  Nevertheless, for complicated models the
retrospective likelihood can remain computationally prohibitive, so
that use of the prospective approach remains widespread. 

In this paper we observe that these likelihood and Bayesian results
are all consequences of certain properties of independence between
parameters.  In \S~\ref{sec:logistic-ml} we show that the results for
maximum likelihood estimation hold whenever we have a strong meta
Markov model, embodying properties of variation independence in the
parameter space.  In \S~\ref{sec:logistic-bayes} we show that the
corresponding Bayesian result holds when, in addition, we use an
overall prior distribution that is a strong hyper Markov law,
exhibiting analogous probabilistic independence between parameters.
In \S~\ref{sec:shm-laws}, we derive parametric classes of strong hyper
Markov laws that can be used for such an analysis, and show that these
encompass the proper prior laws mentioned above.  These results are
further extended to stratified designs in
\S~\ref{sec:strat-case-contr}.

\section{Notation and definitions}
\label{sec:notation-definitions}

Throughout the paper, $(X,Y)$ will denote a single joint observation
from the specified model, and $(X^{(n)},Y^{(n)})$ a sequence of $n$
such observations; $p$ will denote density with respect to an
appropriate measure, with variables indicating the context.  

We recall the notation and definitions of \citet{dawid1993}.  If
$\theta$ denotes a joint probability distribution for $(X,Y)$, then
$\theta_X$ and $\theta_Y$ will denote the corresponding marginal
distributions of $X$ and $Y$, respectively.  We use $\theta_{Y \mid
  X=x}$ to denote the conditional distribution of $Y$ given $X=x$, and
$\theta_{Y \mid X} = (\theta_{Y \mid X=x}:x \in \mathcal{X})$ to
denote the family of all such conditional distributions, labelled by
$x$; we define $\theta_{X \mid Y=y}$, $\theta_{X \mid Y}$ similarly.

A model is a set $\Theta$ of joint probability distributions $\theta$.
A parameter in this model is a function defined on $\Theta$.  We use
the relation $\phi \simeq \psi$ to denote the existence of a bijective
function between the parameters $\phi$ and $\psi$.  For example, we
have $\theta \simeq (\theta_X,\theta_{Y \mid X}) \simeq
(\theta_Y,\theta_{X \mid Y})$.

For two parameters $\phi$ and $\tau$, we define the conditional range of
$\phi$ given $\tau=t$ to be $\{\phi(\theta) : \theta \in \Theta,\ \tau(\theta)
= t \}$.  We say that $\phi$ is {variation independent} of $\tau$, and write
$\phi \vindep \tau$, when this conditional range is constant for all possible
values $t$ of $\tau$: equivalently, when $(\phi,\tau)$ takes values in a
product space.  In a similar manner we can define the {conditional variation
  independence} $\phi \vindep \tau \mid \psi$ \citep{dawid1993}.

A model is called {strong meta Markov} if
\begin{equation}
  \label{eq:smm}
  \theta_X \vindep \theta_{Y \mid X}, \quad \theta_Y \vindep \theta_{X \mid Y}.
\end{equation}

In a Bayesian setting, we use the term \emph{law} to denote a
probability distribution, over the model $\Theta$, for the parameter
variable $\rtheta$ .  We say that a law $\law$ is \emph{strong hyper
  Markov} if we replace the variation independence of \eqref{eq:smm}
with probabilistic independence, denoted by $\indep$, under $\law$:
\[
\rtheta_X \indep \rtheta_{Y \mid X} , \quad
\rtheta_Y \indep \rtheta_{X \mid Y} \quad [\law] .
\]
A necessary, but not sufficient, condition for a law to be strong
hyper Markov is that its support be a strong meta Markov model.

\section{Maximum likelihood estimation in strong meta Markov models}
\label{sec:logistic-ml}

The saturated model, consisting of all probability distributions on
the product space $\mathcal{X} \times \mathcal{Y}$, is trivially
strong meta Markov.  We now investigate some other meta Markov models.

\begin{example}
  \label{exm:contingency-table}
  Let $\nu_X$ and $\nu_Y$ be measures over $\mathcal{X}$ and
  $\mathcal{Y}$ respectively.  The family of all probability
  distributions which have positive densities with respect to $\nu_X
  \times \nu_Y$ is strong meta Markov.
  
  In particular, if $\mathcal{X}$ and $\mathcal{Y}$ are finite, with
  $\nu_X$ and $\nu_Y$ being counting measures, this is the family of
  2-way $|\mathcal{X}| \times |\mathcal{Y}|$ contingency tables
  without structural zeroes.
\end{example}

\begin{example}
  \label{exm:bivariate-normal}
  Let $\Theta$ be the family of bivariate normal distributions for
  $(X,Y)$:
  \begin{equation*}
    \theta = \dnorm\left(
      \begin{bmatrix}
        \mu_X \\ \mu_Y
      \end{bmatrix}
      ,
      \begin{bmatrix}
        \sigma_{XX} & \sigma_{XY} \\
        \sigma_{XY} & \sigma_{YY}
      \end{bmatrix}
    \right) .
  \end{equation*}
  Then $\theta_X = \dnorm( \mu_X, \sigma_{XX})$ and $\theta_{Y \mid
    X=x} = \dnorm ( \mu_{Y \mid X} + \beta_{Y \mid X} x, \sigma_{Y
    \mid X} )$, where
  \begin{equation*}
    \mu_{Y \mid X} = \mu_Y - \frac{\sigma_{XY} \mu_Y}{\sigma_{XX}} 
    , \quad
    \beta_{Y \mid X} = \frac{\sigma_{XY}}{\sigma_{XX}} 
    , \quad
    \sigma_{Y \mid X} = \sigma_{YY} - \frac{\sigma_{XY}^2}{\sigma_{XX}} .
  \end{equation*}
  It is straightforward to establish that $(\mu_X, \sigma_{XX}) \vindep
  (\mu_{Y \mid X}, \beta_{Y \mid X}, \sigma_{Y \mid X})$, and hence that
  $\rtheta_X \vindep \rtheta_{Y \mid X}$, with parallel results when $X$ and
  $Y$ are interchanged.  Therefore this family is a strong meta Markov model.
  This property extends to higher dimensions.
\end{example}

\begin{definition}
  \label{dfn:lambda}
  Suppose the model $\Theta$ consists of a set of joint distributions $\theta$
  for $(X,Y)$ having positive joint density $p(x,y \mid \theta)$.  The
  odds-ratio parameter $\lambda = \lambda(\theta)$ is defined to be the
  labelled collection
  \begin{equation}
    \label{eq:lambda}
    \left( \frac{ 
        p(x,y \mid \theta) \, p(x',y' \mid \theta) 
      }{
        p(x,y' \mid \theta) \, p(x',y \mid \theta)
      }: x,x' \in \mathcal{X};\, y,y'\in \mathcal{Y} \right).
  \end{equation}
\end{definition}

As an example, in the bivariate normal model elements of \eqref{eq:lambda} are
of the form $\exp\left\{- \Lambda_{XY}(x-x')(y-y')\right\}$, where
$\Lambda_{XY} = - \sigma_{XY} / (\sigma_{XX} \sigma_{YY} - \sigma_{XY}^2)$ is
the off-diagonal term of the precision matrix.  Therefore $\lambda \simeq
\Lambda_{XY}$.

The parameter $\lambda$ has been well studied in the context of
contingency tables.  \citet{altham1970} demonstrated that it has
certain desirable properties as a measure of association between $X$
and $Y$.  We note that $\lambda$ also characterises such dependence
for more general models.
\begin{lemma}
  \label{lem:lamba-indep}
  For a given joint distribution $\theta$, $\lambda(\theta) \equiv 1$ if and
  only if $X$ and $Y$ are independent under $\theta$.
\end{lemma}
\begin{proof}
  Now $\lambda \equiv 1$ if and only if
  \begin{equation}
    \label{eq:lambda-indep-1}
    p(x,y \mid \theta) \, p(x',y' \mid \theta)
    =
    p(x,y' \mid \theta) \, p(x',y \mid \theta),
  \end{equation}
  for all $x,y,x',y'$.  If \eqref{eq:lambda-indep-1} holds, then on
  integrating over $x'$ and $y'$ we obtain $p(x,y \mid \theta) = p(x
  \mid \theta) \, p(y \mid \theta)$.  Conversely, if $p(x,y \mid
  \theta)$ factorizes in this manner, \eqref{eq:lambda-indep-1} must
  hold.
\end{proof}

Our particular interest in $\lambda$ is due to its being a common
parameter of both the prospective and retrospective models.
\begin{lemma}
  The odds-ratio $\lambda$ can be expressed as a function of $\theta_{Y \mid X}$, and
  also of $\theta_{X \mid Y}$.
\end{lemma}
\begin{proof}
  Elements of \eqref{eq:lambda} can be written as
  \begin{equation*}
    \frac{
      p(y \mid x, \theta_{Y \mid X}) \, p(y' \mid x', \theta_{Y \mid X})
    }{
      p(y' \mid x, \theta_{Y \mid X}) \, p(y \mid x', \theta_{Y \mid X})
    }
    =
    \frac{
      p(x \mid y, \theta_{X \mid Y}) \, p(x' \mid y', \theta_{X \mid Y})
    }{
      p(x \mid y', \theta_{X \mid Y}) \, p(x' \mid y, \theta_{X \mid Y})
    }  
    .
  \end{equation*}
\end{proof}
As we shall see below, it is this shared parameter property that makes
it possible to use retrospective data to make inferences about the
prospective model.

By constraining $\lambda$, we can construct new strong meta Markov
models:
\begin{lemma}
  \label{lem:constrained-lambda}
  Let $\Theta$ be a strong meta Markov model for $(X, Y)$, and for a given
  function $f$ define $\Theta' = \{ \theta \in \Theta : f (\lambda) = 0 \}$.
  Then $\Theta'$ is strong meta Markov.
\end{lemma}
\begin{proof}
  Since $\theta_{Y \mid X} \vindep \theta_X$ and $f(\lambda)$ is a
  function of $\theta_{Y \mid X}$, it follows from the separoid
  properties of variation independence \citep{dawid2001,dawid2001b}
  that
  \begin{math}
    \theta_{Y \mid X} \vindep \theta_X \mid f(\lambda).
  \end{math}
  Similarly, 
  \begin{math}
    \theta_{X \mid Y} \vindep \theta_Y \mid f(\lambda).
  \end{math}
\end{proof}
\begin{example}
  \label{exm:logistic}
  Let $\mathcal{Y} = \{0,1\}$, and let $\mathcal{X}$ be a subset of
  $\R^d$ whose affine span is $\R^d$.  Let the model $\Theta$ comprise
  all distributions with positive densities on $\mathcal{X} \times
  \mathcal{Y}$.  By the affine condition, there exist $x_1, \ldots ,
  x_{d+1} \in \mathcal{X}$ such that
  $(1,x_1),\ldots,(1,x_{d+1})$ are linearly independent.  We
  can then write $\theta_{Y \mid X} \simeq (\alpha, \beta, \eta)$,
  where
  \begin{equation*}
    p(y \mid x, \alpha, \beta, \eta) =  \frac{e^{y(\alpha + \beta^\top x + \eta_x)}}{1+e^{\alpha + \beta^\top x + \eta_x}},
  \end{equation*}
  with $\eta_x = 0$ for $x = x_1,\ldots,x_{d+1}$.  The odds-ratios are
  then
  \begin{equation*}
    \frac{
      p(1 \mid x, \alpha, \beta, \eta) p(0 \mid x', \alpha, \beta, \eta)
    }{
      p(1 \mid x', \alpha, \beta, \eta) p(0 \mid x, \alpha, \beta, \eta)
    }
    =
    e^{\beta^\top (x-x') + \eta_x - \eta_{x'}}
  \end{equation*}
  and hence $\lambda \simeq (\beta, \eta)$.  The logistic model is
  then obtained on constraining $\eta = 0$.  As $\eta$ is a function
  of $\lambda$, it follows from Lemma~\ref{lem:constrained-lambda}
  that it is strong meta Markov.  Moreover, $\lambda \simeq \beta$ in
  this model.
\end{example}
\begin{example}
  \label{exm:multinomial-logistic}
  We can generalise to let $\mathcal{Y}$ be a finite set.  Applying
  essentially the same argument yields the multinomial logistic model:
  \begin{equation*}
    p(y \mid x, \alpha, \beta) = 
    \begin{cases} 
      \displaystyle \frac{\exp (\alpha_y + \beta_y^\top x) }{ 1+
        \sum_{y' \ne y^*} \exp (\alpha_{y'} + \beta_{y'}^\top x)},
      & \quad y \ne y^* \\
      \displaystyle \frac{1}{ 1+ \sum_{y' \ne y^*} \exp (\alpha_{y'} +
        \beta_{y'}^\top x) }, & \quad y = y^*
    \end{cases}
  \end{equation*}
  for some reference element $y^* \in \mathcal{Y}$.  We then have
  $\lambda \simeq \beta = (\beta_y : y \ne y^*)$.
\end{example}

The cumulative logit model \citep{mccullagh1980}, which is widely used
for ordinal data, is not strong meta Markov.  However there is an
alternative model that can be used in this
 setting:
\begin{example}
  \label{exm:stereotype}
  The \emph{stereotype model} \citep{anderson1984} is obtained by
  constraining the multinomial logistic model so that $\beta_y = \beta
  \gamma_y $, where $\beta \in \R^d$ and $\gamma_y \in \R$.  Then
  $\lambda \simeq (\beta, \gamma)$.  This model can be made more
  general by allowing $\beta$ to take values in $\R^{d \times k}$, and
  $\gamma_y$ in $\R^k$, where $k < |\mathcal{Y}| - 1$.  Several
  authors have proposed this model for ordinal data; in particular
  \citet{greenland1994} noted its validity for analysing retrospective
  data, as we demonstrate below.
\end{example}

\begin{example}
  \label{exm:multi-int}
  The \emph{multiplicative intercept model}
  \citep{hsieh1985,weinberg1993} is a general strong meta Markov model
  for binary response data. It has density of the form
  \begin{equation*}
    p(y \mid x, \alpha, \beta) = \frac{ 
      \bigl\{ e^{\alpha + f(x, \beta)} \bigr\}^y
    }{
      1+ e^{\alpha + f(x, \beta)} 
    }.
  \end{equation*}
  This model can be obtained by constraining the odds-ratios
  \eqref{eq:lambda} to be of the form $f(x, \beta) - f(x', \beta)$.
  It has $\lambda \simeq \beta$.
\end{example}

For the logistic model, \citet{prentice1979} showed that the maximum
likelihood odds-ratio estimators obtained from a case-control study
have the same values and asymptotic distribution as those arising from
a prospective study.  The following result shows that this property
holds for any strong meta Markov model.

\begin{theorem}
  \label{thm:logistic-prof-lik}
  Let $\Theta$ be a strong meta Markov model for $(X,Y)$.  Then the
  profile likelihood function for any function of $\lambda$ is the
  same, up to proportionality, under the joint model $\Theta$, the
  retrospective model $\Theta_{X \mid Y}$ and the prospective model
  $\Theta_{Y \mid X}$.
\end{theorem}
\begin{proof}
  The argument is similar to that of \citet[Lemma 4.10]{dawid1993}.  The joint
  density under the model $\theta$ can be written as $p(x,y \mid \theta) = p(x
  \mid \theta_X) p(y \mid x, \theta_{Y \mid X})$. Therefore the profile
  likelihood $ L^{\text{joint}}_{\mathrm{p}}(\lambda)$ for the joint model is
  \begin{equation}
    \label{eq:logistic-prof-lik-2}
    L^{\text{joint}}_{\mathrm{p}} (\lambda)
    = 
    \max_{\theta : \lambda(\theta) = \lambda} 
    p(x \mid \theta_X) p(y \mid x,\theta_{Y \mid X}).
  \end{equation}
  Since we have the conditional variation independence $\theta_X
  \vindep \theta_{Y \mid X} \mid \lambda$, the maximization in
  \eqref{eq:logistic-prof-lik-2} can be performed separately for each
  factor, hence
  \begin{equation*}
    L^{\text{joint}}_{\mathrm{p}} (\lambda)
    = 
    \max_{\theta_X : \lambda(\theta_X) = \lambda} 
    p(x \mid \theta_X)
    \times \max_{\theta_{Y \mid X} : \lambda(\theta_{Y \mid X}) = \lambda} 
    p(y \mid x,\theta_{Y \mid X}).
  \end{equation*}
  Moreover, since $\theta_X \vindep \theta_{Y \mid X}$ and $\lambda$
  is a function of $\theta_{Y \mid X}$, we have $\theta_X \vindep
  \lambda$, so that the first term is constant for all $\lambda$,
  giving
  \begin{equation*}
    L^{\text{joint}}_{\mathrm{p}} (\lambda)
    \propto 
    \max_{\theta_{Y \mid X} : \lambda (\theta_{Y \mid X}) = \lambda } 
    p(y \mid x,\theta_{Y \mid X}) 
    = 
    L^{\text{pro}}_{\mathrm{p}}(\lambda),
  \end{equation*}
  where $L^{\text{pro}}_{\mathrm{p}}$ denotes the profile likelihood
  of the prospective model.  An identical argument shows that
  \begin{math}
    L^{\text{joint}}_{\mathrm{p}} (\lambda) \propto
    L^{\text{ret}}_{\mathrm{p}}(\lambda).
  \end{math}
  This argument can be extended to any function of $\lambda$.
\end{proof}

From this we obtain the following result, generalizing that of
\citet{prentice1979}.
\begin{corollary}
  \label{cor:logistic-mle}
  Suppose $\Theta$ is a strong meta Markov model parametrized by a
  finite-dimensional parameter.  Then for data observed under
  retrospective sampling, the maximum likelihood estimator of any
  function of the parameter $\lambda$, and its asymptotic covariance
  matrix, can be computed as if the data were observed prospectively.
\end{corollary}
\begin{proof}
  The maximum likelihood estimator is a function of the profile
  likelihood, as is its asymptotic covariance matrix when $\theta$ is
  finite-dimensional \citep{patefield1985}.
\end{proof}
We emphasize that it is necessary for this result that the parameter
of interest be a function of $\lambda$: it is not sufficient that it
be variation independent of the marginals.  In the bivariate normal
example, the correlation coefficient $\rho = \sigma_{XY} /
(\sigma_{XX} \sigma_{YY})^{1/2}$ is variation independent both of
$\theta_X$ and of $\theta_Y$, but cannot be expressed as a function of
either $\theta_{Y \mid X}$ or $\theta_{X \mid Y}$, and cannot be
estimated from a regression.

The above argument can also be applied to the value, but not the
covariance matrix, of a penalized maximum likelihood estimator of
$\lambda$, when the penalty term is a function of $\lambda$ only: for
example, for estimating $\beta$ in a logistic regression by maximizing
$\log p(y \mid x, \alpha, \beta) - \phi(\beta)$ over $\alpha$ and
$\beta$.  Examples of such estimators include ridge regression, where
$\phi(\beta) \propto \|\beta\|_2$, and \lasso, where $\phi(\beta)
\propto \|\beta\|_1$.  Such methods have proven successful in
genome-wide association studies, which involve case-control data with
extremely high-dimensional covariates \citep{park2008,wu2009}. 
\section{Bayesian analysis of retrospective studies}
\label{sec:logistic-bayes}

We now extend the results of the previous section to Bayesian
analysis.  Let $\law$ be a prior law for the parameter variable
$\rtheta\in\Theta$, and let $\law_{\text{pro}}$, $\law_{\text{ret}}$
denote the induced marginal priors for $\rtheta_{Y \mid X}$,
$\rtheta_{X \mid Y}$, respectively.  For observations $(X^{(n)},
Y^{(n)}) = (x^{(n)}, y^{(n)})$, we denote by $\law^{\text{joint}}$ the
posterior law for $\rtheta$, based on prior ${\law}$ and the joint
likelihood $p(x^{(n)}, y^{(n)} \mid \theta)$; by $\law^{\text{pro}}$
the posterior law for $\rtheta_{Y \mid X}$, based on the prior law
$\law_{\text{pro}}$ and the prospective likelihood $p(y^{(n)} \mid
x^{(n)}, \theta_{Y \mid X})$; and by $\law^{\text{ret}}$ the posterior
law for $\rtheta_{X \mid Y}$, based on the prior law
$\law_{\text{ret}}$ and the retrospective likelihood $p(x^{(n)} \mid
y^{(n)}, \theta_{X \mid Y})$.

We now present the key result of this section.
\begin{theorem}
  \label{thm:shm-pro-ret}
  Let $\law$ be a strong hyper Markov prior law over for the joint
  model $\Theta$ for $(X,Y)$.  Then the posterior marginal law of
  $\rlambda = \lambda(\rtheta)$ is the same, whether computed from
  $\law^{\text{\rm joint}}$, from $\law^{\text{\rm pro}}$, or from
  $\law^{\text{\rm ret}}$.
\end{theorem}
\begin{proof}
  The posterior law for $\rlambda$ under the joint analysis is
  determined by its Radon--Nikodym derivative with respect to the
  prior law:
  \begin{equation}
    \label{eq:shm-pro-ret-1}
    \D[\law^{\text{joint}}]{\law} (\lambda)
    \propto
    \int
    \prod_{i=1}^n p( y_i \mid x_i , \theta_{Y|X} ) p(x_i \mid \theta_X) \,\di
    \law(\theta \mid \lambda) .
  \end{equation}
  By the strong hyper Markov property, $\rtheta_{Y|X} \indep \rtheta_X
  \mid \rlambda$, so the right-hand side of \eqref{eq:shm-pro-ret-1}
  factorizes as
  \begin{equation*}
    \int
    \prod_{i=1}^n p( y_i \mid x_i , \theta_{Y|X} ) \,\di
    \law(\theta_{Y|X} \mid \lambda) \ 
    \int
    \prod_{i=1}^n p(x_i \mid \theta_X) \,\di
    \law(\theta_X \mid \lambda) .     
  \end{equation*}
  Also $\rtheta_X \indep \rlambda$, so only the first of these terms
  is a function of $\lambda$.  Therefore
  \begin{equation*}
    \D[\law^{\text{joint}}]{\law} (\lambda)
    \propto
    \int
    \prod_{i=1}^n p( y_i \mid x_i , \theta_{Y|X} ) \,\di
    \law(\theta_{Y|X} \mid \lambda) 
    \propto
    \D[\law^{\text{pro}}]{\law_{\text{pro}}} (\lambda).
  \end{equation*}
  Since the distribution of $\rlambda$ is the same under the priors
  $\law$ and $\law_{\text{pro}}$, the posteriors for $\rlambda$ under
  $\law^{\text{joint}}$ and $\law^{\text{pro}}$ are proportional, and
  hence identical.  A parallel argument shows the identity of the
  joint and the retrospective analyses.
\end{proof}

Several authors have obtained similar results.  \citet{muller1997}
almost identified these conditions for the logistic regression model,
but then incorrectly claimed that the ``argument about the
retrospective likelihood only carries over to posterior inference on
$\beta$ if $\alpha$ and $\beta$ are independent and $\theta_X$ is not
otherwise constrained.''  This misconception appears to be due to the
fact that, although there is a one-to-one mapping between $\alpha$ and
$\theta_Y$, this mapping is itself dependent on $\beta$.
Unfortunately, this means that their proposed Dirichlet process
mixture law does not satisfy the required properties.

For the case of the logistic regression model where the covariate
space $\mathcal{X}$ is finite, conditions equivalent to the strong
hyper Markov property were shown to be sufficient in a 2007 University
of Bristol technical report by A.-M.  Staicu.

The converse result to Theorem~\ref{thm:shm-pro-ret} does not strictly
hold.  For instance, if $\rlambda$ is almost surely constant under the
prior law, then so must it be under any of the posterior laws,
irrespective of whether or not the strong hyper Markov property holds.
However, we conjecture that, with the addition of suitable technical
conditions to exclude such special cases, the identity of the joint,
prospective and retrospective analyses for $\rlambda$ will hold only
when the joint prior law for $\rtheta$ is strong hyper Markov.

It follows immediately from Theorem~\ref{thm:shm-pro-ret} that, with
the stated conditions and definitions, the posterior for $\rlambda$ we
would obtain by combining the true retrospective likelihood with the
prior law $\law_{\text{ret}}$ for its parameter $\rtheta_{X \mid Y}$
could also be obtained by combining the incorrect prospective
likelihood with prior law $\law_{\text{pro}}$ for its parameter
$\rtheta_{Y \mid X}$.  Here we wish to emphasize a constraint that
previous authors have not always made clear: in order to invoke this
result, we must be using a prior law $\law_{\text{ret}}$ for the
retrospective parameter $\rtheta_{X \mid Y}$ that can arise as the
marginal of some strong hyper Markov law $\law$ for $\rtheta$.  Only
then is one justified in using instead the prospective likelihood in
conjunction with a suitable prior law for its parameter $\rtheta_{Y
  \mid X}$---which law we can take to be that derived from $\law$.

The problem of model comparison for case-control studies has received
comparatively little attention in the literature, particularly for
Bayesian analyses.  However we can approach it through a result
similar to that of Theorem~\ref{thm:shm-pro-ret}:
\begin{theorem}
  \label{thm:shm-bayes-factor}
  Let $\law_1(\rtheta)$ and $\law_2(\rtheta)$ be strong hyper Markov
  laws whose marginal laws for $\rtheta_X$ are identical, as are those
  for $\rtheta_Y$.  Then the Bayes factors between $\law_1$ and
  $\law_2$ computed under the prospective, retrospective and joint
  likelihoods are all equal.
\end{theorem}
\begin{proof}
  Define a joint law $\law^*$ for $(\widetilde M,\widetilde \theta)$
  such that $\widetilde M$ takes values $1$ and $2$ each with
  probability $1/2$, and, given $\widetilde M = j$, the conditional
  law of $\widetilde \theta$ is $\law_j$.  The strong hyper Markov
  condition implies
  \begin{equation*}
    \rtheta_X \indep \rtheta_{Y|X} \mid \widetilde M
    \quad [\law^*].
  \end{equation*}
  while the condition of the equality of marginals can be expressed as
  \begin{equation*}
    \rtheta_X \indep \widetilde M
    \quad [\law^*],
  \end{equation*}
  These properties are together equivalent to
  \begin{equation*}
    \rtheta_X \indep (\rtheta_{Y \mid X},\widetilde M)
    \quad [\law^*],
  \end{equation*}
  and similarly 
  \begin{equation*}
    \rtheta_Y \indep (\rtheta_{X \mid Y},\widetilde M)
    \quad [\law^*].
  \end{equation*}
  An argument similar to that of Theorem~\ref{thm:shm-pro-ret} now
  shows that the posterior distributions for $\widetilde M$, and hence
  the Bayes factors, must be the same, whether computed using the
  joint, prospective or retrospective analyses.
\end{proof}

\section{Strong hyper Markov laws}
\label{sec:shm-laws}

We now investigate known families of strong hyper Markov laws, and
methods for deriving new families.  As noted in
\S~\ref{sec:notation-definitions}, strong hyper Markov laws only exist
for strong meta Markov models, so we shall focus on the same models
proposed in \S~\ref{sec:logistic-ml}.  

\citet{dawid1993} identified two strong hyper Markov laws.

\begin{example}
  For discrete $X$ and $Y$, the saturated model comprises all
  multinomial distributions, which can be parametrized by their joint
  probabilities $\theta = (\theta_{x,y}:x\in{\cal X}, y\in{\cal Y})$.
  The standard conjugate prior is a Dirichlet law, $\law(\rtheta) =
  \ddirich(a_{xy}:x\in{\cal X}, y\in{\cal Y})$, with hyperparameters
  $a_{xy}>0$, having density proportional to
  \begin{equation*}
    \prod_{x \in \mathcal{X}, y \in \mathcal{Y}} \theta_{xy}^{a_{xy}-1}.
  \end{equation*}
  The posterior is of the same form, with updated hyperparameters
  $a^*_{xy} = a_{xy} + n_{xy}$, where $n_{xy}$ is the number of cases
  having $X=x, Y=y$.

  By the aggregation properties of the Dirichlet \citep[\eg][Lemma
  7.2]{dawid1993},
  \begin{eqnarray*}
    \rtheta_X &\sim&  \ddirich(a_{x+}:x\in{\cal X})\\
    \rtheta_{Y \mid X = x^*} &\sim& \ddirich(a_{x^*y}:y\in{\cal Y})
    \quad\quad\quad(x^*\in{\cal X})
  \end{eqnarray*}
  all independently, where $a_{x+} = \sum_y a_{xy}$; and similarly for
  $\rtheta_Y$ and $\rtheta_{X \mid Y}$.  Thus this law is strong hyper
  Markov.  Because it is continuous, it also works for the restricted
  model without structural zeroes of
  Example~\ref{exm:contingency-table}.
\end{example}
The Dirichlet law has been widely used for the analysis of
case-control studies with a single binary covariate, corresponding to
a $2 \times 2$ table
\citep{zelen1986,nurminen1987,marshall1988,ashby1993}.  The
distribution of the odds-ratio parameter $\rlambda$ has been explored
by \citet{altham1969}.

\begin{example}
  Consider the bivariate normal model of
  Example~\ref{exm:bivariate-normal}, restricted for simplicity to
  have zero means.  The standard conjugate prior is the inverse
  Wishart distribution for the dispersion matrix $\Sigma$, having
  density proportional to
  \begin{equation*}
    | \Sigma |^a \exp\bigl\{ -\tfrac{1}{2} \tr(A \Sigma) \bigr\}.
  \end{equation*}
  Then the posterior is of the same form, with updated hyperparameters
  $a^*$, $A^*$.  The inverse Wishart distribution determines a strong
  hyper Markov law, with similar marginalization properties to those
  of the Dirichlet law \citep[Lemma 7.4]{dawid1993}.  Similar results
  hold for the non-zero means model, where the conjugate
  normal-inverse Wishart distribution determines a strong hyper Markov
  law.
\end{example}

The independence of the odds-ratio $\rlambda$ from each of the
marginal distributions $\rtheta_X$ and $\rtheta_Y$ allows us to
construct further families of strong hyper Markov laws from existing
ones.
\begin{theorem}
  \label{thm:shm-h}
  If $\law$ is a strong hyper Markov law, then any law $\law'$ having
  Radon-Nikodym derivative of the form
  \begin{equation*}
    \D[\law']{\law}(\theta) = h(\lambda)
  \end{equation*}
  is also strong hyper Markov.  Furthermore, the marginal laws for
  $\rtheta_X$ and $\rtheta_Y$ are the same under $\law'$ as under
  $\law$.
\end{theorem}

\begin{proof}
  Let $A$ be an element of the $\sigma$-algebra generated by
  $\rtheta_{Y \mid X}$.  Since $\rtheta_{Y \mid X} \indep \rtheta_X$
  under $\law$,
  \begin{equation*}
    \law'(A \mid \rtheta_X) 
    = E_\law \bigl[ h\{\lambda(\rtheta_{Y \mid X}) \} \, 1_A (\rtheta_{Y \mid X}) \mid
    \rtheta_X \bigr] 
    = E_\law \bigl\{ h(\rlambda) \, 1_A(\rtheta_{Y \mid X}) \bigr\} = \law' (A),
  \end{equation*}
  and hence $\rtheta_{Y \mid X} \indep \rtheta_X$ under $\law'$.
  Similarly, $\rtheta_{X \mid Y} \indep \rtheta_Y$ under $\law'$.

  Now let $B$ be an element of the $\sigma$-algebra generated by
  $\rtheta_X$.  Then
  \begin{equation*}
    \law'(B) = E_\law \bigl[  h\{\lambda(\rtheta_{Y \mid X}) \} 1_B(\rtheta_X)
    \bigr]
    = E_\law \bigl[  h\{\lambda(\rtheta_{Y \mid X}) \} \bigr] E_\law \bigl\{
    1_B(\rtheta_X) \bigr\} = \law(B) ,
  \end{equation*}
  and similarly for $\rtheta_Y$.
\end{proof}

We can also extend the constraint procedure of
Lemma~\ref{lem:constrained-lambda} to construct strong hyper Markov
laws on the resulting submodel $\Theta'$.
\begin{theorem}
  \label{thm:shm-constrained}
  Let $\law(\rtheta)$ be a strong hyper Markov law, and let $f$ be a
  function of $\lambda$.  Then the law $\law'(\rtheta) = \law(\rtheta
  \mid \widetilde f = 0)$ is strong hyper Markov for the submodel
  $\Theta'$ specified by $f=0$.  Furthermore, the marginal laws for
  $\rtheta_X$ and $\rtheta_Y$ are the same under $\law'$ as under
  $\law$.
\end{theorem}
\begin{proof}
  As $\rtheta_X \indep \rtheta_{Y \mid X}$ and $\widetilde f$ is a
  function of $\rtheta_{Y \mid X}$, we have
  \begin{align}
    \label{eq:m1}
    \rtheta_X \indep \rtheta_{Y \mid X} \mid \widetilde f 
    \quad [\law], \\
    \label{eq:m2}
    \rtheta_X \indep  \widetilde f
    \quad [\law].
  \end{align}
  Parallel results hold with $X$ and $Y$ interchanged.  Then
  (\ref{eq:m1}) shows that $\law(\rtheta)$ remains strong hyper Markov
  under conditioning on $\widetilde f = 0$, while (\ref{eq:m2}) shows
  that this conditioning does not affect the marginal laws.
\end{proof}

\begin{remark}
  \label{rem:arbitrary}
  Together, Theorems~\ref{thm:shm-h} and \ref{thm:shm-constrained} can be
  paraphrased as saying that, if $\law$ is a strong hyper Markov law for
  $\rtheta$, and the law $\law'$ has the same conditional distribution for
  $\rtheta$ given $\rlambda$ as $\law$ does, then $\law'$ is strong hyper
  Markov, with unchanged marginal laws for $\rtheta_X$ and $\rtheta_Y$.  In
  particular, this construction allows $\rlambda$ to be assigned any
  distribution whatsoever under $\law'$.
\end{remark}
\begin{example}
  For a 2-way contingency tables, any law with density of the form
  \begin{equation*}
    h\left( 
      \frac{\theta_{xy}\theta_{x'y'}}{\theta_{xy'}\theta_{x'y}}
    \right)_{x,y,x',y'}
    \prod_{(x,y)} \theta_{xy}^{a_{xy}-1}
  \end{equation*}
  will be strong hyper Markov.  \citet[equation 10]{geiger1997} noted
  that all strong hyper Markov laws for $2 \times 2$ tables must have
  a density of this form.
\end{example}
\begin{example}
  For the zero-means bivariate normal model, any law with density of
  the form
  \begin{equation*}
    h\left(\frac{\sigma_{XY}}{\sigma_{XX}\sigma_{YY} - \sigma_{XY}^2}
    \right)
    | \Sigma |^a \exp\bigl\{ -\tfrac{1}{2} \tr(A \Sigma) \bigr\}
  \end{equation*}
  will be strong hyper Markov.  \citet[Theorem 12]{geiger2002} showed
  that \emph{all} strong hyper Markov laws for the bivariate normal
  must have a density of this form.
\end{example}

The construction of laws for nested models by conditioning on specific
parameters has been proposed by \citet[section 4]{dawid2001a}.
Laws constructed by this procedure will also satisfy the conditions of
Theorem~\ref{thm:shm-bayes-factor}.

\begin{example}
  Consider a logistic model for finite covariate space $\mathcal{X}$,
  as generated by the conditioning procedure of
  Example~\ref{exm:logistic}.

  We start with a generalized Dirichlet law $\law(\rtheta)$ for the
  saturated model.  Then the law for $\rtheta_{Y \mid X}$ has density
  of the form
  \begin{equation*}
    h(\lambda)
    \prod_{x \in \mathcal{X}} \theta_{0 \mid x}^{a_{x0} -1} \theta_{1 \mid x}^{a_{x1} -1}.
  \end{equation*}

  The Jacobian determinant of the transformation to the logistic
  parametrization is
  \begin{equation*}
    \left| \D[\theta_{Y \mid X}]{(\alpha,\beta,\eta)}\right| 
    \propto 
    \prod_{x \in \mathcal{X}} 
    \frac{e^{\alpha + \beta^\top x + \eta_x}}{(1+e^{\alpha + \beta^\top x + \eta_x})^2},
  \end{equation*}
  and hence the density for $\law(\ralpha,\rbeta,\reta)$ is of the
  form
  \begin{equation*}
    g(\beta, \eta)
    \prod_{x \in \mathcal{X}}  \frac{
      e^{(\alpha + \beta^\top x + \eta_x)a_{x1}}
    }{
      (1+e^{\alpha + \beta^\top x + \eta_x})^{a_{x+}}
    },
  \end{equation*}
  where $a_{x+} = a_{x0} + a_{x1}$.  By conditioning on $\reta = 0$,
  we obtain the density of $\law'(\ralpha,\rbeta)$, of the form
  \begin{equation}
    \label{eq:ycondx-dens}
    g(\beta)
    \prod_{x \in \mathcal{X}}  \frac{
      e^{(\alpha + \beta^\top x )a_{x1}}
    }{
      (1+e^{\alpha + \beta^\top x})^{a_{x+}}
    }.
  \end{equation}

  The Jacobian of the transformation in terms of the retrospective
  parameters is
  \begin{equation*}
    \left| \D[(\alpha,\beta,\theta_X)]{(\theta_{X \mid 0},\beta,\theta_{Y=1})} \right| =
    \frac{(1-\theta_{Y=1})^{|\mathcal{X}|-1}}{\theta_{Y=1}} \prod_{x \in \mathcal{X}}
    (1+e^{\alpha+\beta^\top x}).
  \end{equation*}
  Therefore, using a prior law with density \eqref{eq:ycondx-dens} for
  the prospective analysis of retrospective data is justified when the
  true retrospective prior law is
  \begin{equation}
    \label{eq:xcondy-dens}
    g(\beta) \frac{
      \displaystyle \prod_{x \in \mathcal{X}} \theta_{x \mid 0} ^{a_{x+} -1} e^{a_{x1}\beta^\top x}
    }{
      \left(\sum_{x \in \mathcal{X}} \theta_{x \mid 0}  e^{\beta^\top x} \right)^{a_{+1}}
    }.
  \end{equation}
\end{example}

Priors of this form have previously appeared in the literature.  The
prior of \citet[Example 2]{staicu2010} is obtained on rewriting
\eqref{eq:ycondx-dens} as
\begin{equation*}
  g^*(\beta) e^{\alpha a_{+1}} 
  \prod_{x \in \mathcal{X}} \bigl( 1+e^{\alpha + \beta^\top x} \bigr)^{- a_{x+}},
\end{equation*}
where $g^*(\beta) = g(\beta) \exp \bigl( \sum_{x \in \mathcal{X}}
a_{x1} \beta^\top x \bigr)$.  The improper prior of \citet{seaman2004}
and \citet[Example 1]{staicu2010} can be obtained by further taking
the limit as $a_{+1} \to 0$.  However, we argue that the form of
\eqref{eq:ycondx-dens} is more easily interpreted: it can be thought
of as the product of an improper prior with density element $g(\beta)
\, d\beta\, d\alpha$, and a logistic likelihood function, where the
$(a_{xy})$ represent pseudo-counts.  This has the further benefit of
being able easily to adapt existing computational methods: for
example, a Laplace approximation can be found using standard logistic
regression software.

Although $x$ appears in the density (\ref{eq:ycondx-dens}), we
disagree with \citet{staicu2010} that this constitutes a
covariate-dependent prior, like the $g$-priors of \citet{zellner1986}:
it is only dependent on the {a priori} expected frequencies of the
covariates, not on their observed frequencies in the data.

The logistic generalized Dirichlet law can similarly be extended to
the multinomial model of Example~\ref{exm:multinomial-logistic},
yielding density of the form
\begin{equation}
  \label{eq:shm-multinomial}
  g(\beta)
  \prod_{x \in \mathcal{X}}  \frac{
    \prod_{y \ne y^*} e^{(\alpha_y + \beta_y^\top x )a_{xy}}
  }{
    \bigl( 1+ \sum_{y \ne y^*} e^{\alpha_y + \beta_y^\top x} \bigr)^{a_{x+}}
  }.
\end{equation}
By further conditioning this can be applied to the stereotype model of
Example~\ref{exm:stereotype}, using a prior density of the form
\begin{equation}
  \label{eq:shm-stereotype}
  g(\beta,\gamma)
  \prod_{x \in \mathcal{X}}  \frac{
    \prod_{y \ne y^*} e^{(\alpha_y + \gamma_y \beta^\top x )a_{xy}}
  }{
    \bigl( 1+ \sum_{y \ne y^*} e^{\alpha_y + \gamma_y \beta^\top x} \bigr)^{a_{x+}}
  }.
\end{equation}
An analogous construction for the multiplicative-intercept model of
Example~\ref{exm:multi-int} uses a prior density of the form
\begin{equation}
  \label{eq:shm-multi-int}
  g(\beta)
  \prod_{x \in \mathcal{X}}  \frac{
    e^{\{\alpha + f(x,\beta)\} a_{x1}}
  }{
    \bigl\{ 1+ e^{\alpha + f(x,\beta)} \bigr\}^{a_{x+}}
  }.   
\end{equation}

The improper priors of \citet[Theorem 1]{ghosh2012} can be obtained
from \eqref{eq:shm-multinomial}, \eqref{eq:shm-stereotype} and
\eqref{eq:shm-multi-int} by taking the limit $a_{x+} \to 0$.  However
their claim that these priors can also be used for link functions
other than the logistic, such as the probit, skew-symmetric or
cumulative logit, is incorrect, as these models are not strong meta
Markov, and hence can not support strong hyper Markov laws.

The form of the generalized logistic Dirichlet law allows for easy
implementation in generic Bayesian MCMC packages such as {\sc WinBUGS}, {\sc
  OpenBUGS} and {\sc JAGS}, which accept non-integer values for binomial
counts.  Furthermore, arbitrary functions $g$ can be included by use of the
zero Poisson trick: see
\citet[\S~9.5]{bugsbook}.  Unfortunately, this method is somewhat
impractical for large numbers of covariates, since the size of
$\mathcal{X}$ increases exponentially with its dimensionality $k$.
Furthermore, as $\mathcal{X}$ increases, $\rbeta$ will tend to
concentrate around 0.  To compensate for this, the values of
$(a_{xy})$ can be chosen closer to 0, but the above software packages
do not work well for very small values.

\section{Stratified models}
\label{sec:strat-case-contr}

A more complicated analysis is that of stratified or matched
case-control studies, in which participants are selected by both the
outcome $Y$ and an additional stratum variable $S$, taking values in
${\cal S}$.  Such a design can often estimate the odds-ratio of
interest with much greater efficiency than an unstratified study.

It is enough to consider sampling schemes that condition on $S$, so
that the parameter of the joint likelihood is $\theta_{XY \mid S}$.
The prospective parameter of interest is $\theta_{Y \mid XS}$, but
data may be observed under the retrospective regime, only allowing
estimation of $\theta_{X \mid YS}$.  In this case the parameter
$\lambda$ that is a function both of $\theta_{Y \mid XS}$ and of
$\theta_{X \mid YS}$ is the set of all odds-ratios of the form
\begin{equation*}
  \frac{
    p(x,y \mid s,\theta) \, p(x',y' \mid s,\theta) 
  }{
    p(x,y' \mid s,\theta) \, p(x',y \mid s,\theta)  
  }\quad\quad\quad\quad(x,x'\in{\cal X};y,y'\in{\cal Y};s\in{\cal S})
  .
\end{equation*}
\begin{example}
  \label{exm:logistic-strata}
  The stratified logistic model is similar to
  Example~\ref{exm:logistic}, but with an intercept parameter that
  varies by stratum, so that the prospective model is
  \begin{equation*}
    p(y \mid x,s,\alpha,\beta) = \frac{
      e^{\alpha_s + \beta^\top x}
    }{
      1+e^{\alpha_s + \beta^\top x}
    }
    .
  \end{equation*}
  As in the unstratified case, $\lambda \simeq \beta$.
\end{example}

This additional complication can make estimation more difficult.  The
number of strata will typically increase with sample size, with the
result that the maximum likelihood estimator is inconsistent.  An
alternative under the classical approach is to maximize the
conditional likelihood
\begin{equation*}
  L_{\mathrm{c}} (\beta )
  =
  \prod_{s \in \mathcal{S}} \frac{
    \prod_{i \in I_s} e^{y_i \beta^\top x_x}
  }{
    \sum_{\rho} \prod_{i \in I_s} e^{y_{\rho(i)} \beta^\top x_x}
  } ,
\end{equation*}
where $I_s = \{i : s_i = s\}$, and the summation in the denominator is
over the possible permutations of $(y_i)_{i \in I_s}$.  If there are
$a$ cases and $b$ controls in each stratum, called $a{:}b$ matching,
the sum in the denominator will have $(a+b)!/(a!\, b!)$ terms.  In order
to keep this computationally tractable, most studies use $1$:$1$ or
$1$:$m$ matching.

The conditional likelihood does not have a direct Bayesian
interpretation.  \citet[Theorem 1]{rice2004} showed there exists a law
such that the marginal retrospective likelihood $\bar p(x \mid y, s ,
\beta)$ is proportional to the conditional likelihood; however this
law depends on the matching scheme: \eg a $1$:$1$ matched design and a
$1$:$2$ matched design will require different laws.

Alternatively, Theorem~\ref{thm:shm-pro-ret} can be extended to
support use of the prospective likelihood:
\begin{theorem}
  \label{thm:strata-pro-ret}
  Let $\law$ be a prior law for the parameter $\rtheta_{XY \mid S}$ of
  a stratified model, with the property that
  \begin{equation*}
    \rtheta_{Y \mid XS} \indep \rtheta_{X \mid S} , \quad
    \rtheta_{X \mid YS} \indep \rtheta_{Y \mid S}
    \quad [\law] .
  \end{equation*}
  Then the posterior marginal law for the odds-ratios $\rlambda$ is
  the same under the prospective, the retrospective and the joint
  likelihoods.
\end{theorem}
The argument is essentially the same as that for
Theorem~\ref{thm:shm-pro-ret}.

Laws satisfying Theorem~\ref{thm:strata-pro-ret} can be constructed
from a collection of strong hyper Markov laws $\law_s(\rtheta_{XY \mid
  S=s})$ on the individual strata.  A simple example is the
\emph{product law}
\begin{equation*}
  \law (\rtheta_{XY \mid S}) = \prod_{s \in \mathcal{S}} \law_s (\rtheta_{XY \mid S=s}),
\end{equation*}
which is equivalent to fitting a separate model for each stratum, each
having its individual odds-ratio parameter.  The opposite case is that
of a law $\law$ that constrains $\rtheta_{XY \mid S=s} =
\rtheta_{XY \mid S=s'}$ almost surely, thus ignoring stratification
altogether.  However, neither of these extreme cases is able to
exploit the key advantage of stratification, which allows for fitting
a model with both common and stratum-specific parameters, such as the
logistic model in Example~\ref{exm:logistic-strata}, where all strata
share a common odds-ratio.  This can be effected as follows.

\begin{theorem}
  \label{thm:strata-shm} 
  Let $\{\law_s(\rtheta_{XY \mid S=s}):s\in{\cal S}\}$ be a collection of
  strong hyper Markov laws such that the marginal laws for the odds-ratios are
  equal: that is,
  \begin{equation}
    \label{eq:strata-lambda-margin}
    \law_s(\rlambda_s) = \law_{s'}(\rlambda_{s'})
  \end{equation}
  for all $s,s' \in \mathcal{S}$. Then there exists a unique joint
  law $\law(\rtheta_{XY\mid S})$ such that $\law(\rtheta_{XY \mid S=s})
  = \law_s(\rtheta_{XY \mid S=s})$, $\rlambda_s = \rlambda_{s'}$ almost surely,
  and the $(\rtheta_{XY \mid S=s}: s\in{\cal S})$ are conditionally
  independent given $\rlambda$.  Moreover, this law satisfies the
  conditions of Theorem~\ref{thm:strata-pro-ret}.
\end{theorem}

\begin{proof}
  The existence and uniqueness of $\law$ are given by the \emph{Markov
    combination} construction of \citet[Lemma 2.5]{dawid1993}. It
  remains to show that the conditions of
  Theorem~\ref{thm:strata-pro-ret} are satisfied for $\law$.

  The mutual independence of all the $(\rtheta_{XY \mid S=s})$
  conditional on $\rlambda$, combined with the strong hyper Markov
  properties of the $(\law_s)$, implies the mutual independence, given
  $\rlambda$, of all terms of the form $\rtheta_{Y \mid X, S=s}$,
  $\rtheta_{X \mid S=s'}$.  In particular,
  \begin{align}
    \label{eq:strata-shm-3}    
    \rtheta_{Y \mid XS} \indep \rtheta_{X \mid S} \mid \rlambda , \\
    \label{eq:strata-shm-1}
    \bigindep_{s \in \mathcal{S}} \{ \rtheta_{X \mid S
      = s}\} \bigmid \rlambda.
  \end{align}
  Also, since $\law_s$ is strong hyper Markov, we have, for each $s$,
  \begin{equation}
    \label{eq:strata-shm-4}
    \rtheta_{X \mid S = s} \indep \rlambda.
  \end{equation}
  An easy application of the rules of conditional independence shows that
  \eqref{eq:strata-shm-1} and \eqref{eq:strata-shm-4} together imply
  $\rtheta_{X \mid S} \indep \rlambda$, which combined with
  \eqref{eq:strata-shm-3} gives $\rtheta_{Y \mid XS} \indep \rtheta_{X \mid
    S}$, since $\rlambda$ is a function of $\rtheta_{Y \mid XS}$.  Similarly,
  $\rtheta_{X \mid YS} \indep \rtheta_{Y \mid S}$.
\end{proof}

\begin{example}
  \label{exm:logistic-conditional-strata}
  For the stratified logistic model in
  Example~\ref{exm:logistic-strata}, suppose that each law $\law_s$ is
  specified by a density for $(\ralpha_s,\rbeta)$ of the form
  \begin{equation*}
    g_s(\beta) 
    \prod_{x \in \mathcal{X}}  \frac{
      e^{(\alpha_s + \beta^\top x )a_{x1s}}
    }{
      (1+e^{\alpha_s + \beta^\top x})^{a_{x+s}}
    },
  \end{equation*}
  such that the marginal density for $\rbeta$ is $p(\beta)$ in each
  stratum $s$.  By Theorem~\ref{thm:shm-constrained}, this can be
  achieved by choosing
  \begin{equation*}
    g_s(\beta) = \frac{p(\beta)}{
      \int_\R
      \prod_{x \in \mathcal{X}}  \frac{
        e^{(\alpha_s + \beta^\top x )a_{x1s}}
      }{
        (1+e^{\alpha_s + \beta^\top x})^{a_{x+s}}
      } 
      \,\di \alpha_s
    }
    .
  \end{equation*}
  The corresponding joint density for $(\ralpha,\rbeta)$ is then
  \begin{equation*}
    g(\beta) 
    \prod_{(x,s) \in \mathcal{X} \times \mathcal{S}}  \frac{
      e^{(\alpha_s + \beta^\top x )a_{x1s}}
    }{
      (1+e^{\alpha_s + \beta^\top x})^{a_{x+s}}
    }
    \quad \text{where} \quad
    g(\beta) = \frac{\prod_{s \in \mathcal{S}} g_s(\beta)}{p(\beta)^{|\mathcal{S}|-1}}.
  \end{equation*}
  This is of the same form as the density \eqref{eq:ycondx-dens}, where the
  strata are treated as an additional categorical covariate in the model.  As
  with the unmatched case, the improper laws of \citet{ghosh2006,ghosh2012}
  can be obtained by taking the limit $a_{xys} \to 0$, though again the claims
  in \citet{ghosh2012} regarding the use of different link functions are
  incorrect.  Similar priors can be obtained for the multinomial and
  stereotype models in the previous section.

  Again, we emphasize that using such a law for the prospective
  analysis of retrospective data requires that the prior law
  $\law(\rtheta_{X \mid YS})$ be the marginal of a joint law such that
  $\rtheta_{Y \mid XS} \indep \rtheta_{X \mid S}$ and $\law(\rtheta_{X
    \mid S=s}) = \ddirich(a_{xs})$.
\end{example}

We have not specified a model for the stratum variable $S$, as we have
assumed all data are observed conditional on $S$.  However, under the
additional assumption $\rtheta_{XY \mid S} \indep \rtheta_S \  [\law]$, 
the data can be treated as if they were randomly sampled from the
population, as would hold for a cross-sectional study.

\section{Discussion}
\label{sec:discussion}

We have outlined a broad framework with necessary assumptions for the
analysis of retrospective data using a prospective likelihood or
Bayesian approach.

Our Bayesian analysis requires the existence of a joint strong hyper
Markov law of which the prospective and retrospective laws are its
margins.  Because of the difficulties of defining and handling
marginalization for improper priors \citep{dawid1973}, our arguments
do not readily extend to improper priors, whose use in this context
may require a different justification.

These results only apply to functions of the odds-ratio.  Other
quantities such as an intercept parameter $\alpha$ cannot be inferred
using this approach, nor does it incorporate more recent developments
such as case-cohort designs and incorporation of population incidence
data.

Many analyses \citep[\eg][]{devocht2012} have used multivariate normal prior
laws for the logistic log odds-parameter $\rbeta \simeq \rlambda$; but the
overall laws used are not strong hyper Markov, and the resulting prospective
and retrospective posterior laws for $\rbeta$ are not equal.  However,
Remark~\ref{rem:arbitrary} shows that it is indeed possible to construct a
strong hyper Markov law such that $\rbeta$ is multivariate normal; and the
previously suggested prior laws might possibly be interpretable as
approximating such a strong hyper Markov law.  There could nevertheless be
considerable difficulty in determining the precise form of the implied law for
the retrospective parameters.

Similar properties and techniques arise in other contexts. A recent example is
the development of inverse regression techniques used for dimension
reduction \citep{cook2009,taddy2013}. These methods exploit the existence of
low-dimensional representations of the odds-ratio $\lambda$, termed a
sufficient reduction, and utilise a similar method of obtaining estimates by
fitting the wrong inverse model to the data.

Another example arises in the computation of graphical \lasso
estimators for high-dimensional covariance matrices
\citep{banerjee2008,friedman2008}. These are shrinkage estimators which
penalize off-diagonal elements of the precision matrix. Due to the strong meta
Markov property of the multivariate normal model and the penalized terms being
functions of the odds ratio, a similar argument to
Theorem~\ref{thm:logistic-prof-lik} can be used to show that the solution to
the optimisation problem is equivalent to a set of penalized regression
problems of each covariate against all others. As a result, the estimate can
be computed by an iterative scheme of \lasso regressions.

\section*{Acknowledgements}
The research of the first author is supported by an EPSRC Postdoctoral Fellowship.

\bibliographystyle{biometrika} \bibliography{refs}
\end{document}